\documentclass[fleqn,twoside]{article}
\usepackage{espcrc2}

\readRCS
 $Id: espcrc2.tex,v 1.2 2004/02/24 11:22:11 spepping Exp $
\usepackage{graphicx}

\newcommand{\AmS}{{\protect\the\textfont2
  A\kern-.1667em\lower.5ex\hbox{M}\kern-.125emS}}

\title{ Search for deconfinement in the cluster at ultrarelativistic heavy ion collisions}

\author{ M. K. Suleymanov\address[j]{ Department of Physics, \\ 
        COMSATS Institute of Information Technology, Islamabad}%
\thanks{E-mail: mais$\_$suleymanov@comsats.edu.pk},        
        E. U. Khan\addressmark[j],
        K. Ahmed\addressmark[j],
        Mahnaz Q. Haseeb\addressmark[j], 
        Farida Tahir\addressmark[j]
        and
        Y. H. Huseynaliyev\addressmark[j] }

\begin{document}

\begin{abstract}
Some of the centrality experiments indicate regime change and 
saturation in the behavior of characteristics of the secondary 
particles. It is observed as a critical phenomenon for hadron-nuclear, 
nuclear-nuclear interactions and ultrarelativistic heavy ion 
collisions. The existing simple models do not explain the effect. 
We believe that the responsible mechanism to explain the phenomena could be the percolation 
cluster formation and expect appearance of deconfinement in the cluster.  
\vspace{1pc}
\end{abstract}

\maketitle

\section{INTRODUCTION}

Quarkonia production is among prime observables in ultrarelativistic 
heavy ion physics~\cite{[1]} and expected to be suppressed in the 
Quark-Gluon Plasma (QGP). Anomalous suppression of the production of 
$J/\psi$ was observed in central $Pb-Pb$ collisions at SPS CERN. It 
was expected that the effect would be stronger at RHIC BNL~\cite{[2]}. 
The results demonstrated that heavy flavor particles are very sensitive to 
phase transition of strongly interacting matter and to formation of QGP.   

Study of centrality dependence of the characteristics of nuclear-nuclear 
interactions is an important experimental way to obtain information on phases of 
strongly interacting matter formed during the collision evolution. 
If the regime change observed in different experiments takes place unambiguously 
twice, this would be the most direct experimental evidence to a phase transition from 
hadronic matter to a phase of deconfined quarks and gluons. However, the second point of 
regime change has not been observed clearly. 

Fig. 1 shows the experimental ratios of the average values of multiplicity of $K^{+}-, K^{-}-,\phi -$ mesons 
and $\Lambda$-hyperons to the average values of multiplicity of  $\pi^{\pm}$- mesons as a function of 
centrality~\cite{[3]}. The regime change and saturation for the behavior of the ratios as 
a function of centrality can be seen. Similar result was obtained for  pion production and stopping in $proton-Be, Cu,$ 
and $Au$ collisions as a function of centrality at a beam momentum of $18 GeV/c$~\cite{[4]}, for $\Lambda$ 
production as a function of collision centrality  for $p–Au$ collisions at   $17.5 GeV /c$ and for $K^0_s$ and $K^+-$ mesons 
emitted in $p+Au$ reaction at AGS energies. 
\begin{figure}[htb]
\includegraphics[height=15pc,width=20pc]{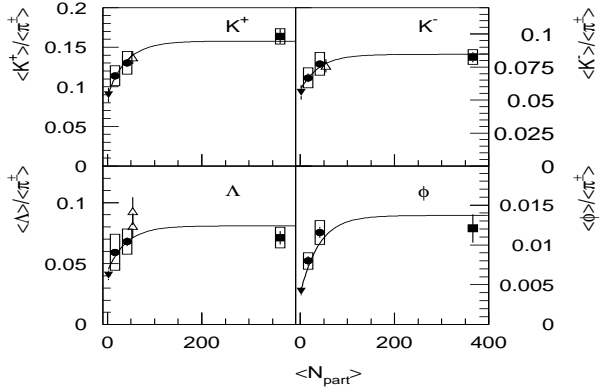}
\caption{The ratio of the average values of multiplicity of strange particles to the ones of pion 
multiplicity as a function of the centrality - $n_{part}$~[3]}
\label{fig:1}
\end{figure}

\section{ REGIME CHANGE AND SATURATION IN CHARMONIUM PRODUCTION}

The regime change has been observed: at some values of centrality~\cite{[5]} and colliding energy, 
as a critical  phenomena; for hadron-nuclear, nuclear-nuclear interactions and 
ultrarelativistic ion collisions; in the ranges of energy from SIS energy up to RHIC 
energy; almost for all particles. After the point of regime change, saturation is 
observed. The existing simple models cannot explain the effect. For this it is necessary 
to suggest that the dynamics is the same for all such interactions, independent of 
energy and mass of the colliding nuclei and their types. The mechanism to describe the phenomena 
may be statistical or percolative due to their critical character ~\cite{[6]}. We believe that 
the mechanism to explain the phenomena could be the percolation cluster formation~\cite{[7]}, which 
makes a multibaryon system. Big percolation clusters may be formed in these interactions independent 
of the colliding energy but the structure, maximum density and temperature of hadronic matter may 
depend on colliding energy and masses in the cluster framework. 
  The deconfinement is expected when the density of quarks and gluons becomes so high that due to 
strong overlap, it no longer makes sense to partition them into color-neutral hadrons~\cite{[8]}. 
The clusters get much larger than hadrons, within which color is not confined; deconfinement is thus 
related to cluster formation and a connection between it and percolation seems very likely~\cite{[9]}. 
The charmonium suppression can also be a result of deconfinement in clusters.   

\section{ APPEARANCE OF CRITICAL TRANSPARENCY IN STRONGLY INTERACTING MATTER }

  The heavy flavor particles are most sensitive to phase transition and to formation of QGP. 
Observance of the effects connected with formation and decay of the percolation clusters in heavy 
ion collisions at ultrarelativistic energies and the study of correlation between these effects and 
charmonium suppression could be a confirmation of deconfinement of strongly interacting matter in clusters. 
Increasing the centrality of collisions, its size and masses 
could increase as well as its absorption capability, we may see the saturation. So after point of regime 
change the conduction of the matter increases and it becomes a superconductor~\cite{[8]}-\cite{[9]} 
due to the formation of 
percolation cluster with the quarks bound, as a result. 
  The critical change of transparency could influence the characteristics of secondary particles.  
Thus to confirm the deconfinement in cluster it is necessary to study the centrality dependence 
of behavior of heavy flavor particles yield and simultaneously, critical increase in the transparency 
of the strongly interacting matter.

\end{document}